\documentclass[preprint,amsmath,amssymb,aps,prstab]{revtex4-2}

\usepackage[utf8]{inputenc}
\usepackage[T1]{fontenc}
\usepackage{textcomp}
\usepackage{gensymb}

\usepackage{graphicx}
\usepackage{dcolumn}
\usepackage{bm}
\usepackage{hyperref}

\usepackage{xspace}
\usepackage{amsmath}
\usepackage{amsfonts}
\usepackage{epstopdf}
\usepackage{epsfig}
\usepackage{fancyhdr}
\usepackage{tabularx}
\usepackage{amssymb}
\usepackage{array,booktabs}
\usepackage{multirow}
\usepackage{babel}
\usepackage{graphics}
\usepackage{graphicx}
\usepackage{subfigure}
\usepackage{comment}

\newcommand{\be}{\begin{equation}}
\newcommand{\ee}{\end{equation}}

\newcommand{\nts}{\negthickspace}
\newcommand{\nms}{\negmedspace}

\newcommand{\amit}{\textsc{AMIT}\xspace}
\newcommand{\opal}{\textsc{OPAL}\xspace}

\begin{document}

\preprint{APS/PRAB}

\title{Beam stripping interactions in compact cyclotrons}

\author{P.~Calvo}
\email{pedro.calvo@ciemat.es}
\author{I.~Podadera}
\author{D.~Gavela}
\author{C.~Oliver}
\affiliation{Centro de Investigaciones Energéticas, Medioambientales y Tecnológicas (CIEMAT), Madrid, Spain.}

\author{A.~Adelmann}
\author{J.~Snuverink}
\author{A.~Gsell}
\affiliation{Paul Scherrer Institut (PSI), Villigen, Switzerland}

\date{\today}


\begin{abstract}

Beam stripping losses of H$^-$ ion beams by interactions with residual gas and electromagnetic fields are evaluated. These processes play an important role in compact cyclotrons where the beam is produced on an internal ion source, and they operate under high magnetic field. The implementation of stripping interactions into the beam dynamics code \opal provides an adequate framework to estimate the stripping losses for compact cyclotrons such as \amit. The analysis is focused on optimizing the high energy beam current delivered to the target. The optimization is performed by adjusting parameters of the ion source to regulate the vacuum level inside the accelerator and minimize the beam stripping losses.


\end{abstract}

\maketitle


\section{Introduction}

Radioisotopes production for medical applications has led to the development and improvement of cyclotrons, which currently play a remarkable role in the sustainability of the radionuclide supply~\cite{oliver-1}, especially for positron emission tomography (PET) scans. The progress on the cyclotron technology over the last decades ~\cite{schmor} has resulted in very mature, compact and versatile products, which are commercially available in the market~\cite{smirnov}. As a result of this progress, the number of cyclotron facilities for medical purposes is expanding worldwide~\cite{IAEA-06}, offering alternatives for the production of medical radioisotopes such as $^{99m}$Tc~\cite{boschi}, traditionally produced in nuclear reactors.

Cyclotrons for biomedical radionuclide production are typically based on the acceleration of light ions up to tens of MeV. Hydrogen anions are selected in most of the present facilities as the accelerated light ion beam because of the simplification of the beam extraction from the cyclotron to the target. For the production of H$^-$ beam, the use of internal ion sources has demonstrated to be the most cost-effective and compact solution~\cite{ehlers,an}. Extraction systems based on thin stripper foils results in an effective mechanism able to modify the extracted ion energy, while avoiding any turn separation requirement at the extraction point~\cite{jasna,kleeven}. In addition, some last-generation cyclotrons integrate superconducting magnets to provide intense magnetic fields with advantages in power consumption, size, weight, and therefore in compactness.

Nevertheless, the use of H$^-$ internal ion sources has as a drawback the deterioration of the vacuum level in the acceleration chamber, especially in the central region~\cite{ehlers,yang}. Part of the H$_2$ gas injected continuously in the ion source to produce the H$^-$ beam is extracted from the ion source slit together with the beam, configuring the main source of residual gas of the vacuum chamber. Since the binding energy of the hydrogen anions ($<\!1$~eV)~\cite{lykke} is low, the interaction between the H$_2$ residual gas and the H$^-$ beam plays an important role in the transport of the beam along the cyclotron. In addition, the high magnetic field produces a strong electric field in the rest-frame of the H$^-$, which could induce the detachment of the second slightly bound electron limiting the lifetime of high-energy ions. These interactions reduce the survival rate of the beam and consequently the total beam transmission. As we will discuss in the following sections, they represent a non-negligible source of sustained and uncontrolled losses that could damage the vacuum chamber walls and/or activate the chamber, increasing the maintenance time and reducing the availability. Therefore, the magnitude of the impact of these issues for compact cyclotrons~\cite{papash,Zhang,pradhan} imposes a dedicated study, which could be also applied to other type of accelerators~\cite{carneiro,plum}.

In this paper, firstly, we review H$^-$ beam stripping interactions from residual gas and magnetic fields. We then describe the implementation of those physical processes into the beam dynamics code \opal~\cite{opal}. And finally, we use this implementation to evaluate the impact and optimize the beam dynamics of the AMIT cyclotron~\cite{cyclomed}, which is a cutting-edge compact accelerator aimed to produce on-demand short-lived radioisotopes to enhance medical nuclear imaging. This is achieved by an autonomous and small-sized design in order to provide an alternative to the centralized radioisotope production.


\section{Beam stripping interactions}

The stripping interactions are stochastic processes in which each individual particle could experience the interaction according to a certain probability. Let's assume incident particles on a homogeneous medium are subjected to a process with a mean free path $\lambda$ between interactions. Hence, the probability of suffering an interaction before reaching a path length $x$ is~\cite{leo}:
\begin{equation}
P(x) \;=\; 1 - e^{-x/\lambda},
\label{eq:prob_1}
\end{equation}
where $P(x)$ represents the cumulative interaction probability of the process.

When the interaction occurs between a beam colliding with particles of a material, the mean free path is described in terms of the cross section, $\sigma$, and the particle density of interacting centers of the material, $n$:
\begin{equation}
\lambda = \frac{1}{n\,\sigma}.
\end{equation}
Likewise, the mean free path could be connected with the mean time of interaction or lifetime, $\tau$:
\begin{equation}
\tau=\frac{\lambda}{\beta c}=\frac{1}{n\sigma\beta c},
\end{equation}
where $\beta$ is the ratio of the velocity of the particles over the speed of light, $c$.


\subsection{Residual gas stripping}
\label{sec:residual}

Residual beam interactions with the molecules of the remaining gas in the acceleration region entail the most important source of losses in a H$^-$ compact cyclotron.
Assuming a beam flux incident in a gas with density $N$ (number of the gas molecules per unit volume under the vacuum condition), the beam fraction lost per unit of traveled length will be, in compliance with Eq.~\eqref{eq:prob_1}:
\begin{equation}
f_g=1-e^{-N\sigma x},
\label{eq:fraction_lost_gas}
\end{equation}
where the gas density, $N$, is easily obtained from the ideal gas equation. If several types of molecules are considered in the residual gas as well as different physical processes of comparable significance, the total mean free path is given by:
\begin{equation}
\frac{1}{\lambda_{total}}=\sum_j \frac{1}{\lambda_j}=N_{total}\cdot\sigma_{total}=\sum_jN_j\,\sigma^{\,j}_{total}=\sum_j\left(\sum_iN_j\,\sigma^{\,j}_{i}\right),
\end{equation}
where $N_j$ is the particle density of each gas component of the residual gas, and $\sigma_i$ represents the cross section of each of the physical phenomena.

In case of H$^-$ ion beams the second electron is bound with a low energy of $\varepsilon=0.754195(19)$~eV~\cite{lykke}. Hence, there is a relevant probability of electron detachment during the acceleration process through interaction with the residual gas. 

The experimental cross section for charge-transfer interactions of hydrogen atoms and ions has been measured since the 1950s for interactions with different gases. The data have been published and compiled in different reports~\cite{Allison,barnett-2,Nakai,phelpsH2,barnett-chebyshev,phelpsAr}. Moreover, theoretical studies for high energy cross sections have been developed based on an extension of the Bethe theory for the total inelastic cross section in the Born approximation~\cite{Gillespie1,Gillespie2,Gillespie3}.

In addition, different analytic methods to fit the cross section data have been formulated. One of them~\cite{barnett-chebyshev} is based on fitting the recommended cross sections as function of the projectile energy, $E$, using least-squares method with Chebyshev polynomials:
\begin{equation}
\ln{[\sigma(E)]}=\frac{1}{2}a_0 + \sum_{i=1}^{k} a_i \cdot T_i(X),
\label{eq:barnett}
\end{equation}
\begin{equation}
X=\frac{(\ln{E}-\ln{E_{min}})-(\ln{E_{max}}-\ln{E})}{\ln{E_{max}}-\ln{E_{min}}},
\end{equation}
\vspace{-2pt}

\noindent where $T_i$ are the Chebyshev orthogonal polynomials, $a_i$ ($i=0,1,...,k$) denote adjustable parameters relative to each reaction, $k$ is the smallest number of coefficients providing an accurate fit and $E_{min}$ and $E_{max}$ are parameters that limit the region of analytic representation of the cross section. This procedure facilitates the data interpolation in a restricted measured energy range; however, it cannot be used for extrapolation, because it often shows non-physical behavior just outside the considered energy interval.

A more robust method~\cite{Nakai} makes use of analytic expressions that approximate low-energy and high-energy asymptotic trends. It is based on a semi-empirical expression for inelastic collision cross sections~\cite{Green} developing functional forms to fit a compiled set of experimental data using the two-step least-squared method. For the considered reactions, the analytical function formula is given by the general expression:
\begin{equation}
\sigma_{qq'} = \sigma_0 \left[ f(E_1) + a_7\!\cdot\!f(E_1/a_8) \right],
\label{eq:CS_Nakai}
\end{equation}
where $qq'$ represents any combination of initial and final charge states of the particle, $\sigma_0$ is a convenient cross section unit ($\sigma_0 = 1\!\cdot\!10^{-16}\;\text{cm}^2$), whereas $f(E)$ and $E_1$ are given by:
\begin{equation}
f(E) = \frac{ a_1\!\cdot\!\left(\!\displaystyle\frac{E}{E_R}\!\right)^{\!a_2} }{ 1+\left(\!\displaystyle\frac{E}{a_3}\!\right)^{\!a_2+a_4}\!\!+\left(\!\displaystyle\frac{E}{a_5}\!\right)^{\!a_2+a_6} },
\label{eq:CS_Nakai2}
\end{equation}
\vspace{-2mm}
\begin{equation}
E_1 = E_0 - E_{th},
\label{eq:CS_Nakai4}
\end{equation}
being $E_0$ the energy of the incident particle in keV, $E_{th}$ is the threshold energy of the reaction in keV, the symbols $a_i$ denote adjustable parameters and $E_R$ is the Rydberg energy multiplied by the ratio of the atomic hydrogen mass, $m_{H}$, to the electron mass, $m_e$:
\begin{equation}
E_R = hcR_{\infty}\!\cdot\!\frac{m_H}{m_e} = \frac{m_He^4}{8\varepsilon_0^2h^2},
\label{eq:CS_Nakai3}
\end{equation}
where $h$ is the Planck constant, $c$ is the speed of light in vacuum, $R_{\infty}$ is the Rydberg constant, $e$ is elementary charge and $\varepsilon_0$ is the vacuum electric permittivity.

This analytic expression has been improved for reactions of hydrogen ions with hydrogen gas~\cite{TabataShi} by means of linear combinations of $f(E)$, taking into account additional experimental data and considering more setting parameters. The enhancement of the function makes it possible to extrapolate the cross section data to some extent.


\subsection{Electromagnetic stripping}
\label{sec:lorentz}

H$^-$ ions traveling in a high-magnetic field might undergo single-electron detachment reactions due to the opposite bending force experienced by electrons and nucleus according to their electric charge. The field component orthogonal to the velocity of the particles produces an electric field, $\mathcal{E}$, in the ion's rest-frame according to the Lorentz transformation ($\mathcal{E}\!=\!\gamma\beta c B$). The strength of this effect is a function of the energy of the ions and the magnetic field, $B$.  This effect, called electromagnetic or Lorentz stripping, is relevant just for H$^-$ beams due to the low binding energy of the second electron. Only single-electron detachment processes are expected~\cite{Furman} because of the high binding energy of the first electron ($13.598434600291(12)$ eV~\cite{Mohr}). The fraction of H$^-$ beam particles dissociated by the electromagnetic field after a traveled distance $L$ can be evaluated through the interaction probability (see Eq.~\eqref{eq:prob_1}):
\begin{equation}
f=1-e^{\,-\,L/\beta c\gamma\tau}=1-e^{\,-\,t/\gamma\tau},
\end{equation}
where $\tau$ is the particle lifetime in the rest frame of the ions.

The process can be analyzed from the decay of an atomic system in a weak and static electric field. For sufficiently high fields, the potential perceived by the electrons is modified, decreasing at some distance below the binding energies. The dissociation of the atomic system occurs by tunneling through the potential barrier into a decay channel giving rise to the ground state of the daughter atom. A theoretical study based on the calculation of the electric dissociation rate directly from the formal theory of decay have obtained an expression for the of H$^-$ ion lifetime~\cite{Scherk}:
\begin{equation}
\tau= \frac{4m_ez_T}{S_0\mathcal{N}^2\hslash\,(1+p)^2\left(1-\displaystyle\frac{1}{2k_0z_t}\right)}\,\cdot\,\exp\!{\left(\frac{4k_0z_T}{3}\right)}
\label{lifetime}
\end{equation}
where $z_T=\varepsilon/eE$ is the outer classical turning radius, $\varepsilon$ is the electron binding energy, $S_0$ is the spectroscopic coefficient of finding the daughter atom in the ground state, $p$ is the polarization of the ionic wave function, $k_0$ is a parameter derived from the ionic wave function and determined from the relation $k_0^2\!=\!2m_e\varepsilon/\hslash^2$ and $\mathcal{N}$ is a normalization factor given by:
\begin{equation}
\mathcal{N}=\frac{[2k_0(k_0+\alpha)(2k_0+\alpha)]^{1/2}}{\alpha},
\end{equation}
being $\alpha$ a parameter for the ionic potential function.

This lifetime expression can be parametrized in terms of the electric field, $\mathcal{E}$ and the atomic properties:
\begin{equation}
\label{lifetime_}
\tau =
\frac{\sqrt{2\varepsilon m_e}\,\alpha^2}{S_0(1+p)^2\,e\left(\nts\displaystyle\frac{\sqrt{2\varepsilon m_e}}{\hslash}\!+\!\alpha\nms\right)\nms\nms\left(\nts\displaystyle\frac{\sqrt{8\varepsilon m_e}}{\hslash}\!+\!\alpha\nms\right)\nms\nms\left(\nts1\!-\!\displaystyle\frac{\sqrt{2}\,e\,\hslash\,\mathcal{E}}{4\sqrt{m_e}\,\varepsilon^{3/2}}\right)}\,\frac{1}{\mathcal{E}}\cdot \exp{\nms\left[\frac{2}{3}\!\left(\frac{\displaystyle8m_e}{\displaystyle e^2\,\hslash^2}\right)^{\nts\nts1/2}\!\frac{\varepsilon^{3/2}}{\mathcal{E}}\right]},
\end{equation}

\noindent where $m_e$ is the electron mass, $e$ is the elementary charge and $\hbar$ is the reduced Planck constant. The values of the atomic parameters are already known: $p~=~0.0126$~\cite{Scherk}, $S_0=0.783\,(5)$~\cite{Keating} and $\alpha=3.806\!\cdot\!10^{10}\;\text{m}^{-1}$~\cite{Tietz}. This expression can be written in a simpler way as:
\begin{equation}
\tau=\frac{a_F}{(1-\eta \mathcal{E})\,\mathcal{E}}\,\cdot\,\exp{\!\left(\frac{b_F}{\mathcal{E}}\right)},
\label{lifetime_P2}
\end{equation}

\begin{equation}
a_F=\frac{\sqrt{2\varepsilon m_e}\,\alpha^2}{S_0\,(1+p)^2\,e\left(\nts\displaystyle\frac{\sqrt{2\varepsilon m_e}}{\hslash}\!+\!\alpha\nms\right)\nms\nms\left(\nts\displaystyle\frac{\sqrt{8\varepsilon m_e}}{\hslash}\!+\!\alpha\nms\right)},
\end{equation}

\begin{equation}
b_F=\frac{2}{3}\left(\frac{\displaystyle8m_e}{\displaystyle e^2\,\hslash^2}\right)^{\nms\nms1/2}\!\varepsilon^{3/2} = \frac{4}{3}\frac{\varepsilon^{3/2}}{\sqrt{e\,\hslash\,\mu_B}},
\end{equation}

\begin{equation}
\eta=\frac{\sqrt{2}\,e\,\hslash}{4\sqrt{m_e}\,\varepsilon^{3/2}} = \sqrt{\frac{m_e}{2}}\,\frac{\mu_B}{\varepsilon^{3/2}},
\end{equation}
\vspace{0.5mm}

\noindent where $\mu_B$ is the Bohr magneton. The parameters depend on fundamental constants and ion properties, taking values of $a_F=2.65\,(2)\cdot 10^{-6}$ s V/m, $b_F=4.4741\,(2)\cdot 10^{9}$ V/m and $\eta=1.49007\,(5)\cdot 10^{-10}$ m/V.

Apart from that, prior to this theoretical approach, a simpler expression for the rest-frame lifetime was obtained experimentally~\cite{Stinson,Jason,Keating} as a function of some setting parameters ($A_1$, $A_2$):
\begin{equation}
\tau=\frac{A_1}{\mathcal{E}}\,\cdot\,\exp{\!\left(\frac{A_2}{\mathcal{E}}\right)}.
\end{equation}
This expression has been widely used in previous studies as a first approximation to determine the influence of the magnetic field in a H$^-$ beam. However, it could be derived from the theoretical lifetime (Eqs.~\eqref{lifetime_} or~\eqref{lifetime_P2}) taking into account some approximations: disregarding the polarization factor, $p$, and considering $\eta E \!<\!<\!1$. Thus, a relationship for the experimental constants $A_1$ and $A_2$ as a function of the binding energy can be found as:
\begin{equation}
A_1(\varepsilon)=C_1\,\frac{\varepsilon}{S_0N^2},
\end{equation}
\vspace{-3mm}
\begin{equation}
A_2(\varepsilon)=C_2\,\varepsilon^{3/2},
\end{equation}
where $C_1$ and $C_2$ are constants independent of the atomic structure that can be referred to as fundamental constants~\cite{nist,codata} by means of the theoretical expression under the considered approximation:
\begin{equation}
C_1=\frac{4\,m_e}{e\,\hslash}=\frac{2}{\mu_B}=2.157\cdot 10^{23} \;\text{A}^{-1}\,\text{m}^{-2},
\end{equation}
\begin{equation}
C_2=\frac{4}{3}\frac{1}{\sqrt{e\,\hslash\,\mu_B}}=1.065\cdot 10^{38}\;\;\text{T}^{1/2}\text{A}^{-1/2}\,\text{J}^{-1}\,\text{s}^{-1}.
\end{equation}
Hence, the parameters of the experimental mean lifetime expression can be reassessed: $A_1=2.714\,(17)\cdot10^{-6}$ s V/m and $A_2=4.47407\,(17)\cdot10^{9}$ V/m, in good agreement with previous values~\cite{Jason,Keating}. It is important to emphasize that these parameters have always been determined by experimental adjustments. Therefore, their assignment based on fundamental constants provides an important enhancement for the study of electromagnetic stripping.

The evaluation of the lifetime for a H$^-$ beam shows that the Lorentz stripping becomes relevant for high-energy beams immersed in high-magnetic fields. Hence, this process has to be assessed and quantified in the design of an accelerator in order to prevent unexpected beam losses.


\section{\opal computational model for beam stripping}

\opal (Object Oriented Parallel Accelerator Library) is a versatile open source Particle-In-Cell (PIC) code developed for large-scale charged-particle optics simulations in accelerators and beam lines~\cite{opal}. It is organized into two different flavors, one of them (\textsc{OPAL-cycl}) exclusively dedicated to cyclotrons. The particles are evolved in time by either a fourth-order Runge-Kutta method or a second-order leapfrog according to the collision-less Vlasov-Poisson equation.

One of the more attractive features of \opal is the capability to perform beam interactions with matter through Monte Carlo simulations. The physical reactions considered by the code have been recently extended with the implementation of beam stripping reactions~\cite{calvo}. This new feature enhances the potential and versatility of OPAL, providing the capability to extend beam dynamics studies to cyclotrons where this could be a relevant issue.

The beam stripping algorithm evaluates the interaction of hydrogen ions with the residual gas and the electromagnetic fields. In the first case, the cross sections of the processes are estimated according to the energy by means of analytical functions (see Sec.~\ref{sec:residual}). The implementation allows the user to set the pressure, temperature and composition of the residual gas, which could be selected for the calculations as either molecular hydrogen (H$_2$) or dry air in the usual proportion~\cite{picard}. For precise simulations, a 2D pressure field map from an external file can be imported into \opal, providing more realistic vacuum conditions. Concerning electromagnetic stripping, the electric dissociation lifetime of H$^-$ ions is evaluated through the theoretical formalism (see Sec.~\ref{sec:lorentz}). In both instances, the individual probability at each integration step for every particle is assessed.

\begin{figure}[h]
	\centering
	\includegraphics [width=0.55\textwidth]  {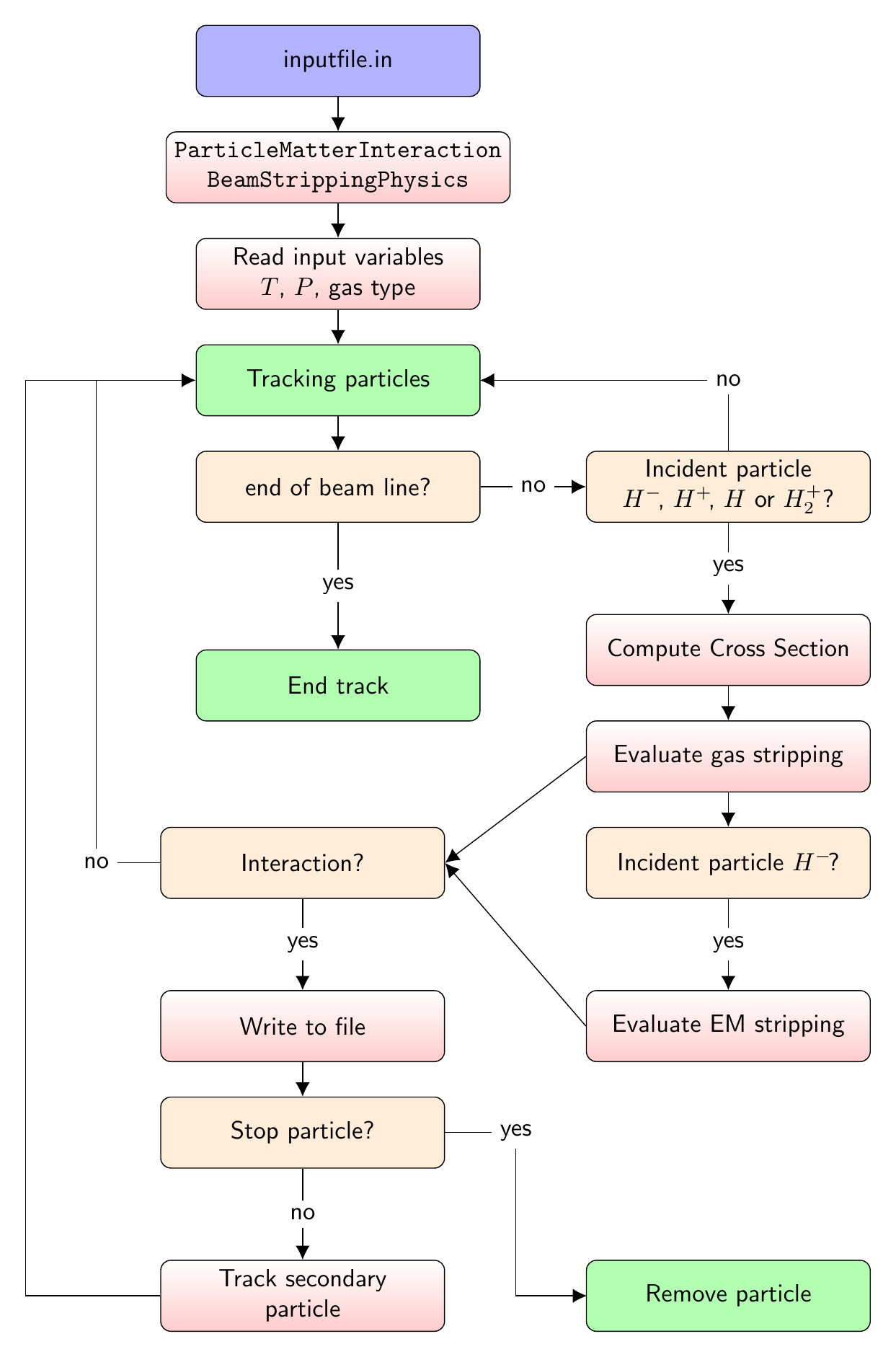}
	\caption{\opal beam stripping physics algorithm flowchart.}
	\label{fig:flowchart}
\end{figure}

A stochastic evaluation method through an uniformly generated random number is used to evaluate if a physical reaction occurs. In case of interaction, it will be stripped and removed from the beam, or optionally transformed to a secondary heavy particle, compliant with the occurred physical phenomena. In this case, the secondary particle will continue its movement in agreement with the charge-to-mass ratio. Fig.~\ref{fig:flowchart} summarizes the iterative steps evaluated by the algorithm until the end of the accelerating process, or until the particle is removed from the beam.

\section{Beam stripping assessment}

\subsection{\amit cyclotron}
\label{sec:amitCyclotron}
\begin{figure}[!t]
	\centering
	\includegraphics[width=8.6cm]{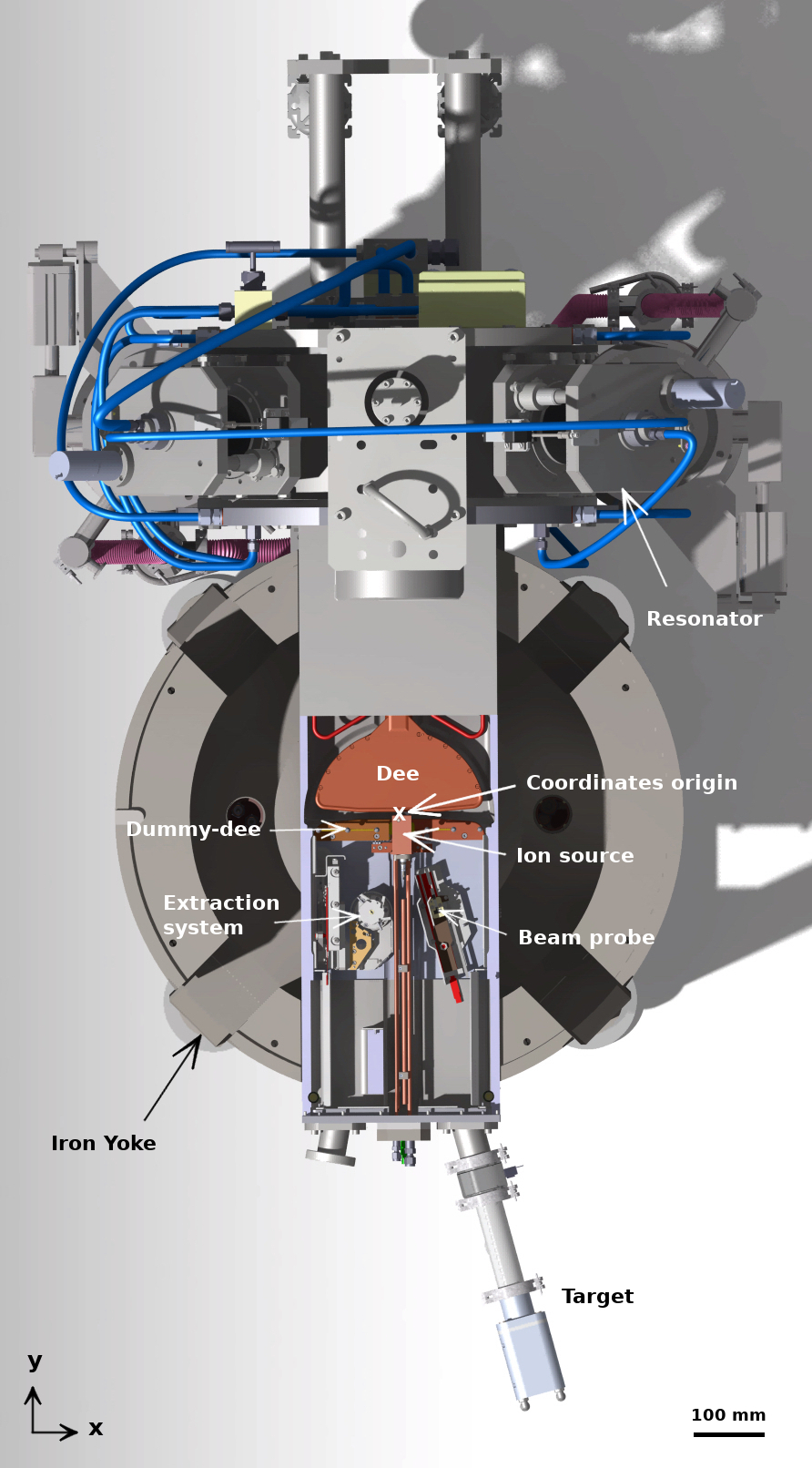}
	\caption{Top view of the \amit cyclotron 3D model. The main components, the center and the axes of the global reference frame and the scale are indicated.}
	\label{fig:AMIT_3D}
\end{figure}

The \amit (Advanced Molecular Imaging Technologies) cyclotron is a superconducting weak focusing cyclotron designed for single-dose production of $^{18}$F and $^{11}$C radionuclides. It is a Lawrence-type machine accelerating a H$^-$ beam current of $10\,\mu$A up to $8.5$~MeV~\cite{oliver}. Its compact configuration (Fig.~\ref{fig:AMIT_3D}) is aimed to contribute to the deployment in hospitals for on-site production of short-life radioisotopes. The very compact design of the accelerator is achieved by the combination of: 1) a high magnetic field (central value of $4$~T) created by a superconducting magnet of NbTi~\cite{garcia-tabares}, and 2) with the use of an internal Penning Ionization Gauge (PIG)-type ion source. A radiofrequency system~\cite{gavela-2} of $60.134$~MHz in a one $180\degree$ dee configuration, at the end of a quarter wave coaxial resonator, provides the electric field for the beam acceleration. The required final energy imposes a $60$~kV accelerating peak voltage. An independent vacuum chamber, adapted to the magnet aperture, holds the accelerating system, the ion source, the extraction system, and the beam diagnostics. The stripping-based extraction provides the final proton beam, making use of a high-efficiency mechanism through thin carbon foils. The evaluation of the extraction process resulted in total charge transfer with low energy losses for the optimized foil thickness~\cite{calvo_phd}. The stripper is installed in a movable mechanical structure, enabling the energy tuning of the resultant beam.


\subsection{Vacuum system}

The vacuum system of the \amit cyclotron has been designed with the goal of achieving at least an absolute residual gas pressure below $10^{-5}$~hPa in the acceleration chamber. The system has to deal with the compactness of the accelerator that limits the space available to install the vacuum pumps considering the fringe magnetic field and the radiation environment among other issues. The independent acceleration chamber of $700$~mm length, $52$~mm height, and $280$~mm width is made of copper (surface area of $0.26$~m$^2$) and steel (surface area of $0.5$~m$^2$). It hosts the internal ion source, aligned with the axial direction, and the RF cavity with a $12$~mm free height inside the dee electrode.

The vacuum system has been simulated with Molflow+~\cite{molflow-1,molflow} to evaluate different configurations of the vacuum pumps, taking into account the pumping system capacity, the impact on the whole accelerator size, the maintenance requirements, as well the interactions with other cyclotron subsystems. From those simulations, a configuration with two diffusion pumps connected to the cylindrical resonator was chosen as a preferred option. The pumps system has the capability to achieve an ultimate pressure below $10^{-7}\!$~hPa, but due to the continuous gas injection from the ion source ($0-10$~sccm), the pressure level in the chamber during operation is much higher (around $10^{-4}-10^{-5}\!$~hPa).

Molflow+ considers steady state molecular flow of an ideal gas for the simulations. Therefore, the mean free path of the residual gas particles or molecules is considerably greater than the dimensions of the vacuum vessel. The vacuum level is determined by the throughput, depending on the pressure gradient and the conductance. Vacuum calculation for the \amit chamber has been performed with Molflow+ considering H$_2$ residual gas, a conductance of $0.70\;\text{l}/\text{s}$ and a throughput of $4.3\cdot10^{-3}\;\text{Pa}\!\cdot\!\text{m}^3/\text{s}\,= 2.546\;\text{sccm}$.
The results are easily scalable if other input parameters have to be considered.

\begin{figure}[!b]
	\centering
	\subfigure[]
	{\includegraphics [width=8.6cm] {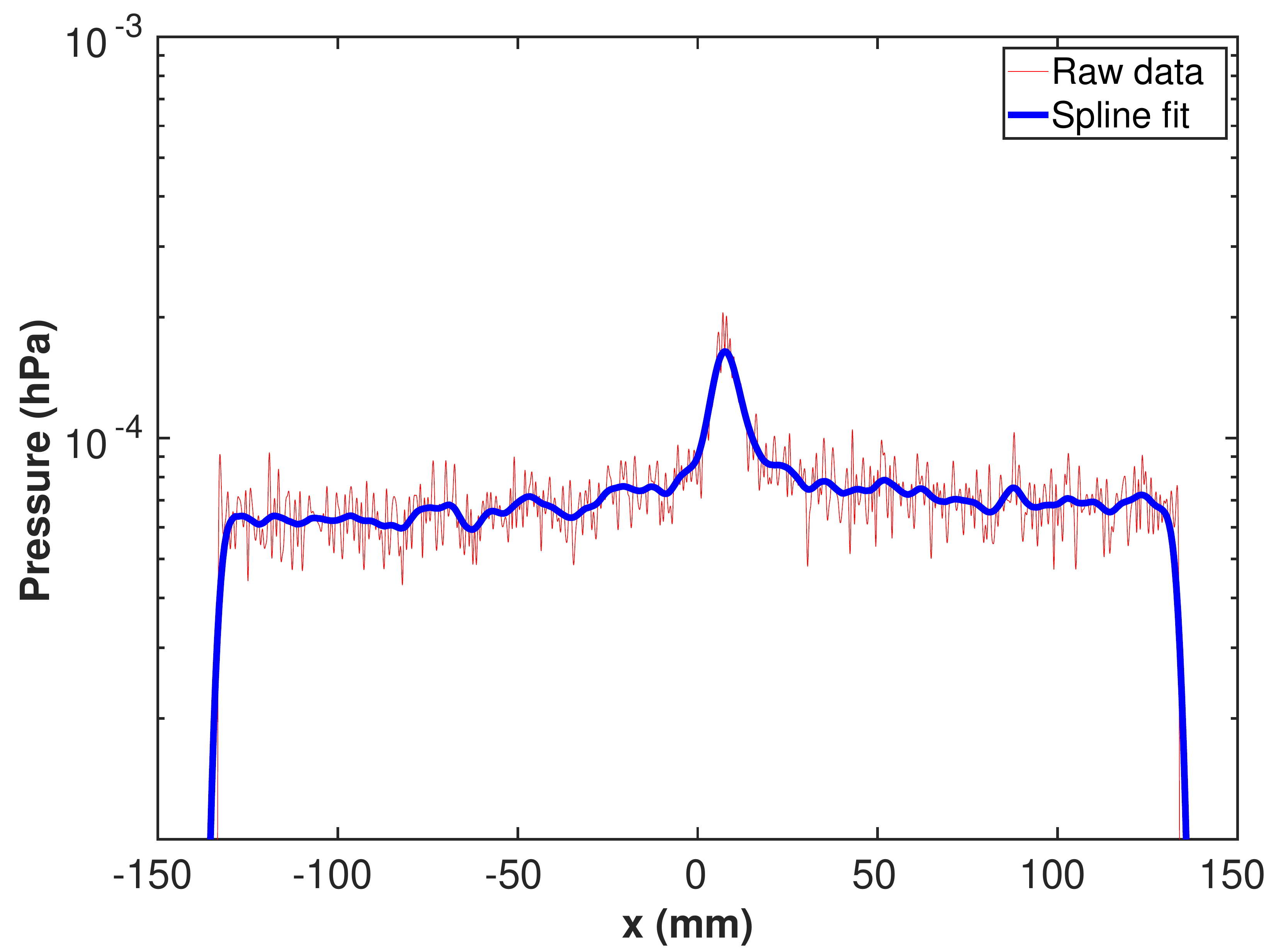}} $\;$
	\subfigure[]
	{\includegraphics[width=8.6cm]{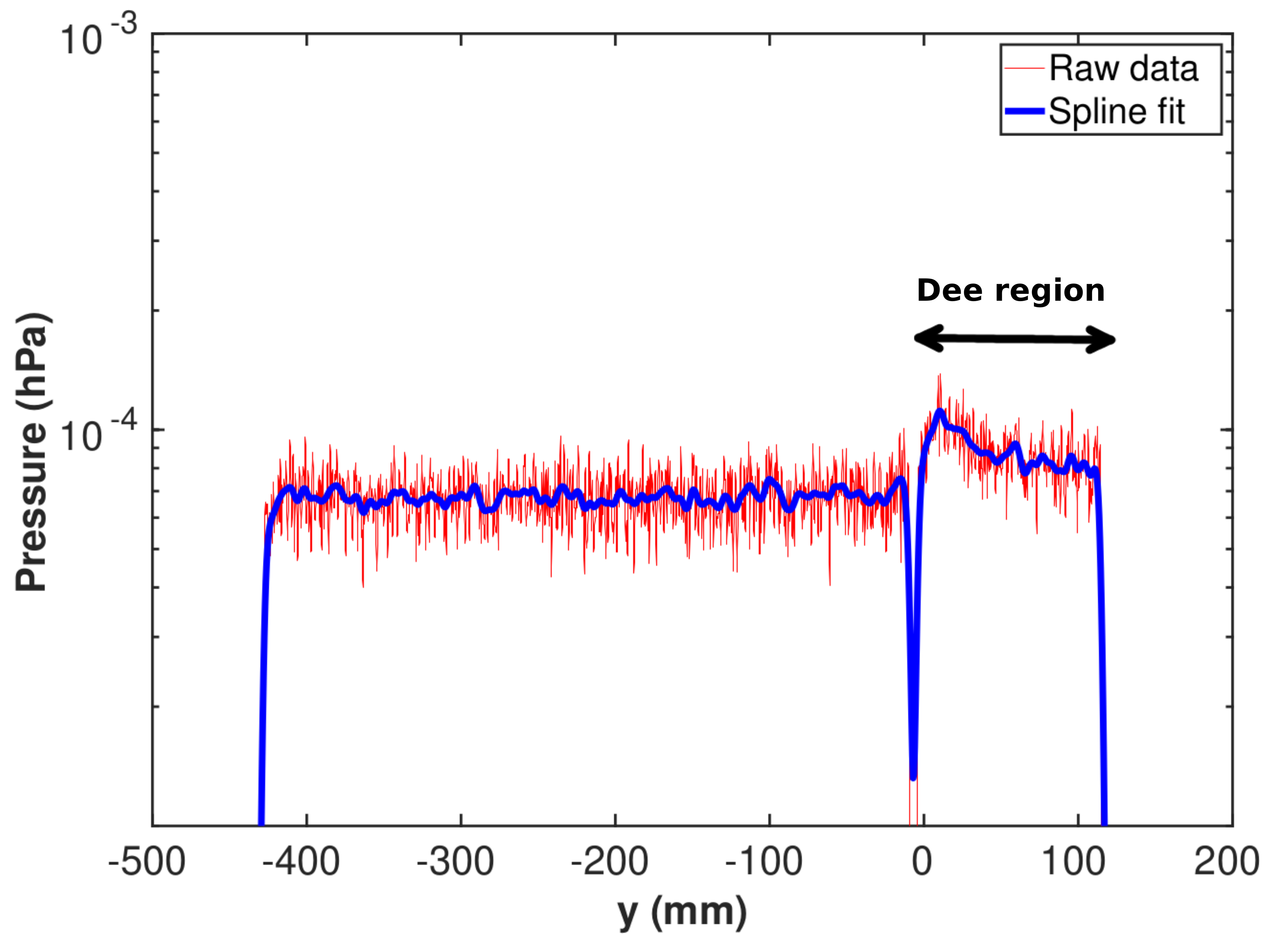}}
	\caption{Pressure level along $x$-coordinate at the gap center (a) and along $y$-coordinate at $x=0$ (b). The figures are referred to the cyclotron center (see Fig.~\ref{fig:AMIT_3D}). Thus, $y$-positive values in subfigure (b) correspond to the dee region, whereas the negative part represents the dummy-dee area and the rest of the vacuum chamber. The red line represents the raw data from Molflow+ simulations and the blue line is a spline fit.}
	\label{fig:PressureFieldmapXY}
\end{figure}

The simulation results (see Fig.~\ref{fig:PressureFieldmapXY}) show an average value of the pressure of $7\cdot10^{-5}$~hPa. However, the vacuum level is worse in the central region ($\sim 1.7\cdot10^{-4}$~hPa), because in front of the ion source slit the gas flux from the source and the low conductance increases the pressure significantly. Furthermore, the dee region has a slightly larger pressure due to the gas evacuation limitations through the opening holes at the back of the structure. Fig.~\ref{fig:PressureFieldmapXY} illustrates the pressure level in the vacuum chamber along the center of the accelerating gap (at $y=0$) and along the perpendicular direction passing through the center of the machine (at $x=0$). Given this vacuum level, a relevant effect of the beam stripping interactions in the beam transmission through the cyclotron is expected.


\subsection{\amit beam stripping analysis}

The compact design of the \amit cyclotron requires a careful analysis of the beam stripping interactions in order to evaluate their contribution to the final beam losses. The implementation of these particle-matter interactions into \opal has allowed the assessment of both, the electromagnetic stripping and the residual gas interactions. The beam dynamics simulations with beam stripping interactions have been performed under the nominal accelerations conditions of the cyclotrons as it has been exposed in the previous section~\ref{sec:amitCyclotron}.

\subsubsection{Electromagnetic stripping}

A preliminary calculation shows that a $4$~T magnetic field for the maximum achievable energy of $8.5$~MeV in the \amit cyclotron corresponds to a beam-rest-frame electric field of $E\!=\!160$~MV/m. This entails a marginal beam fraction loss per unit length of $1.42\cdot10^{-6}$~m$^{-1}$ due to electromagnetic stripping obtained from the theoretical evaluation. Therefore, Lorentz stripping is not expected to have any noteworthy contribution to the stripping losses for the current cyclotron configuration due to the low energy of the beam. The \opal simulations have corroborated the negligible contribution of electromagnetic stripping to the beam losses, and consequently it can be considered as a source of second order losses.

\subsubsection{Gas stripping}

\begin{figure}[!t]
	\centering
	\subfigure[]
	{\includegraphics [width=8.6cm] {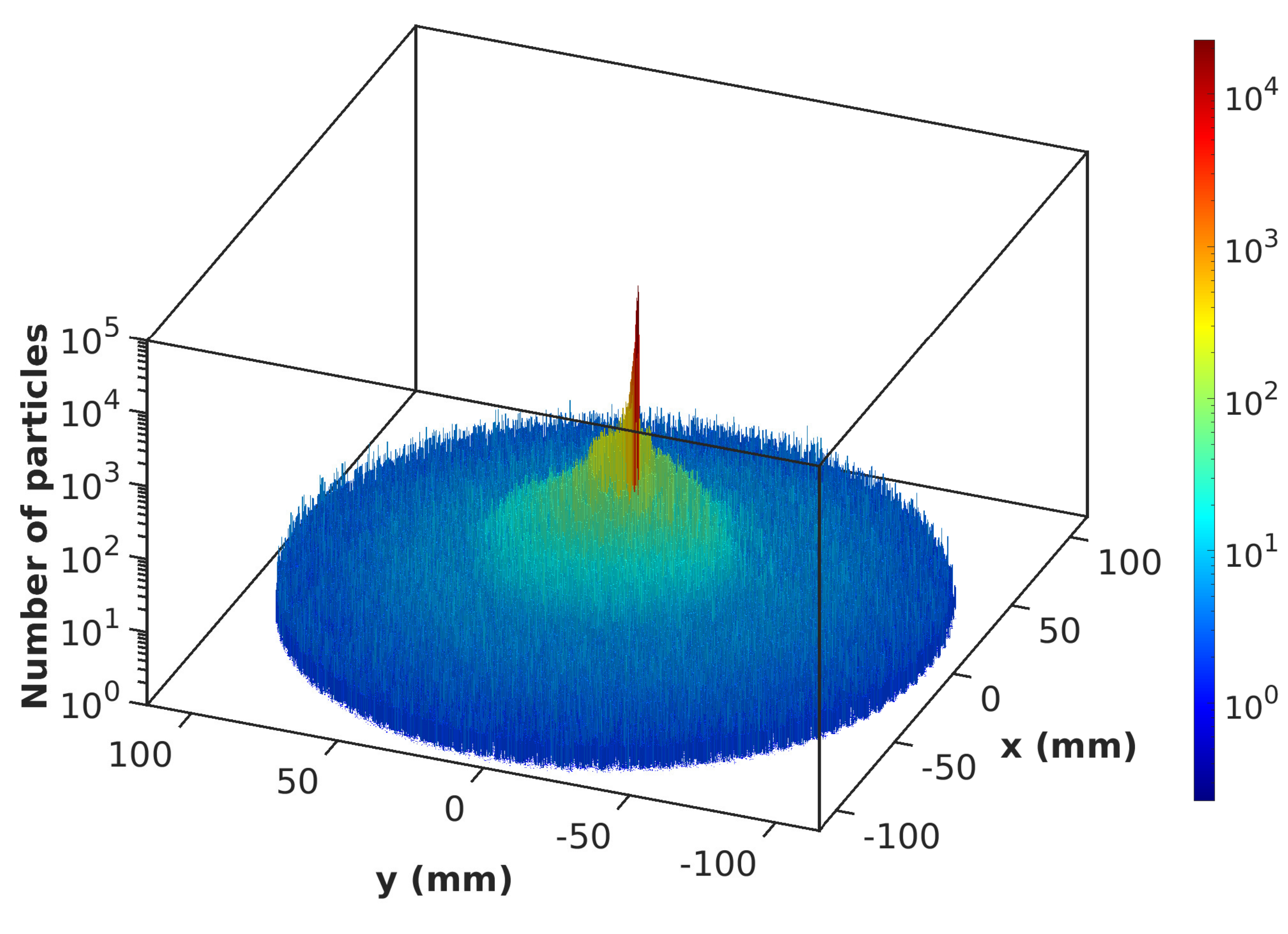}} \qquad
	\subfigure[]
	{\includegraphics [width=8.6cm] {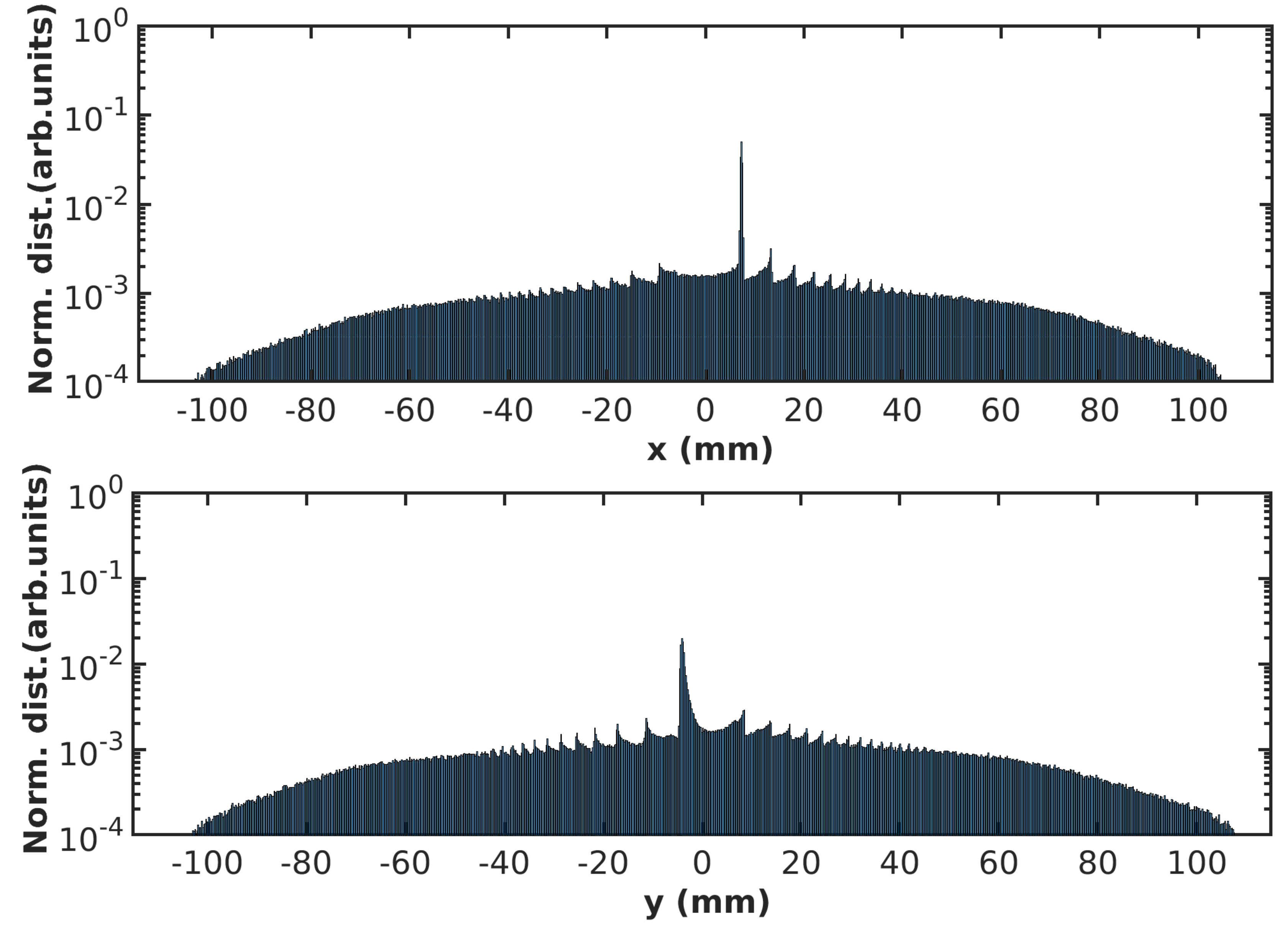}}
	\caption{(a) Spatial distributions in the cyclotron plane of the beam losses produced by gas stripping. (b) Normalized distribution over the total amount of stripping losses along different directions.}
	\label{fig:BSTP-hist_spatial}
\end{figure}

The gas stripping losses along the acceleration process in \amit cyclotron has been characterized through \opal simulations including the pressure field map obtained from Molflow+ results. The average rate of stripping losses estimated by the simulation reaches $(14.6\pm1.5)\,\%$ under the considered vacuum conditions. The uncertainty of the result is due to the stochastic error associated with stripping losses. Fig.~\ref{fig:BSTP-hist_spatial} illustrates the spatial characterization of the beam stripping losses along the cyclotron. They are predominantly produced in the first stages of the acceleration, some millimeters after the ion source, for two main reasons. Firstly, the pressure in the central region is one order of magnitude higher in comparison with the rest of the chamber. Secondly, the cross section of H$^-$ interactions with the gas molecules is maximal for the energy range $(0.1-300)$~keV of the beam in the central region. Nevertheless, the stripping interactions persist along the beam acceleration at a lower level given the long traveled distances until the extraction.

\begin{figure}[!t]
	\centering
	\includegraphics [width=8.6cm] {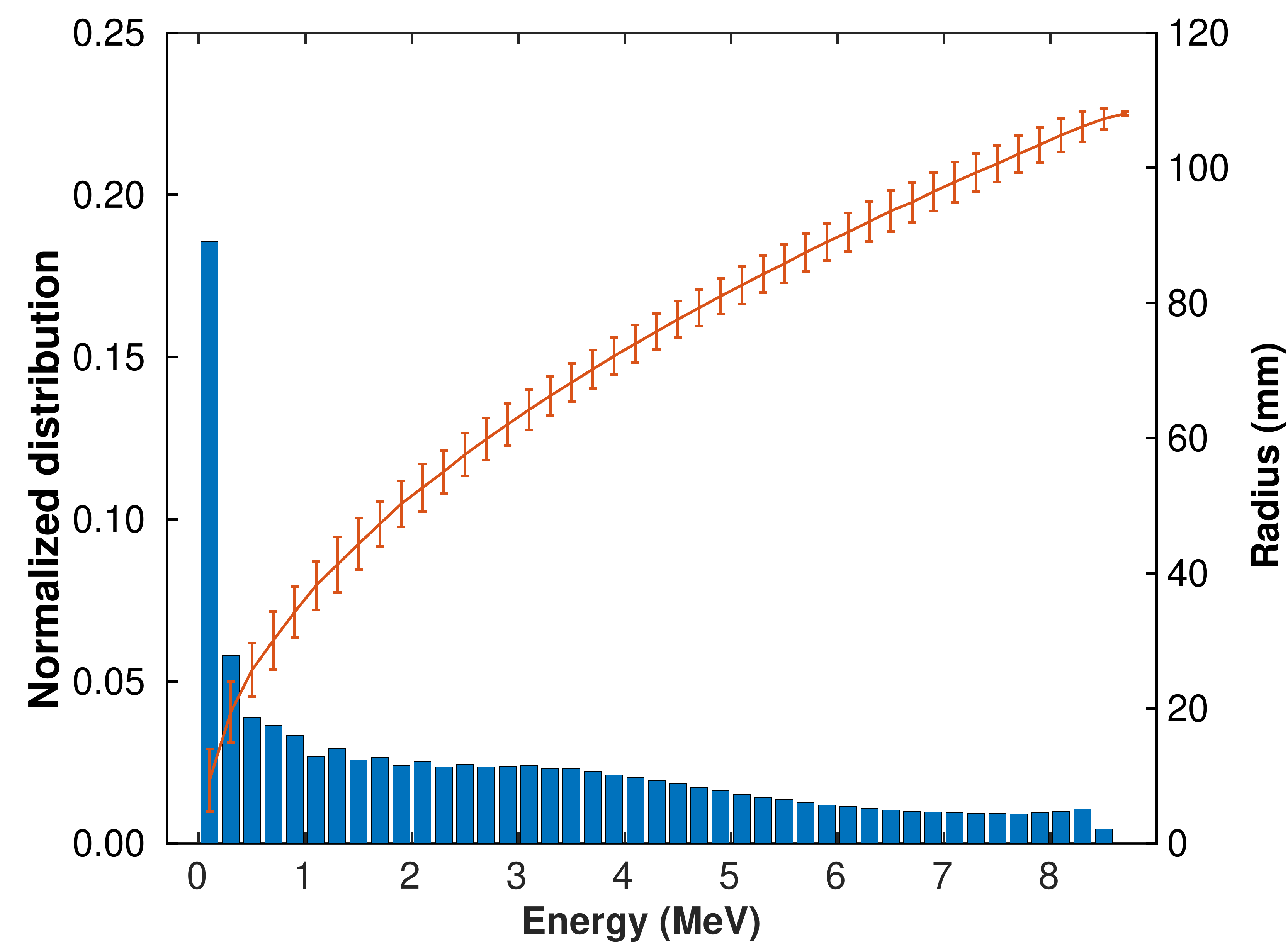}
	\caption{Histogram of normalized energy distribution of the losses. The normalization is over the total amount of stripping losses. Solid orange line with error bars represents the radial location of particles corresponding to each histogram bin associated.}
	\label{fig:BSTP-hist}
\end{figure}

The energy characterization of the beam stripping losses is presented in Fig.~\ref{fig:BSTP-hist}. It is manifest that the main part of stripped particles induces low-energy losses ($<\!1$ MeV). However, the prolongation of the stripping interactions along the entire beam path induces high-energy losses, which will entail the continuous damage and degradation of several components in the long term and the increase of the radioactive activation by secondary particles of energy above $2.5$~MeV.

Given the relevance of the beam stripping reactions caused by the interactions with the residual gas in a compact cyclotron, it is essential to look for some improvements to minimize their impact. The vacuum level in the acceleration chamber is directly related to the amount of neutral gas injected, depending on the input gas flow in the ion source and on the size of the chimney slit. The modifications of these parameters involve a linear variation of the pressure of the vacuum chamber under the consideration of a steady state molecular flow from the source. Thus, they can be adjusted to reduce the stripping losses in the \amit cyclotron as well as to modify the injected current.

\begin{figure}[!b]
    \centering
    \subfigure[]
	{\includegraphics [width=8.6cm] {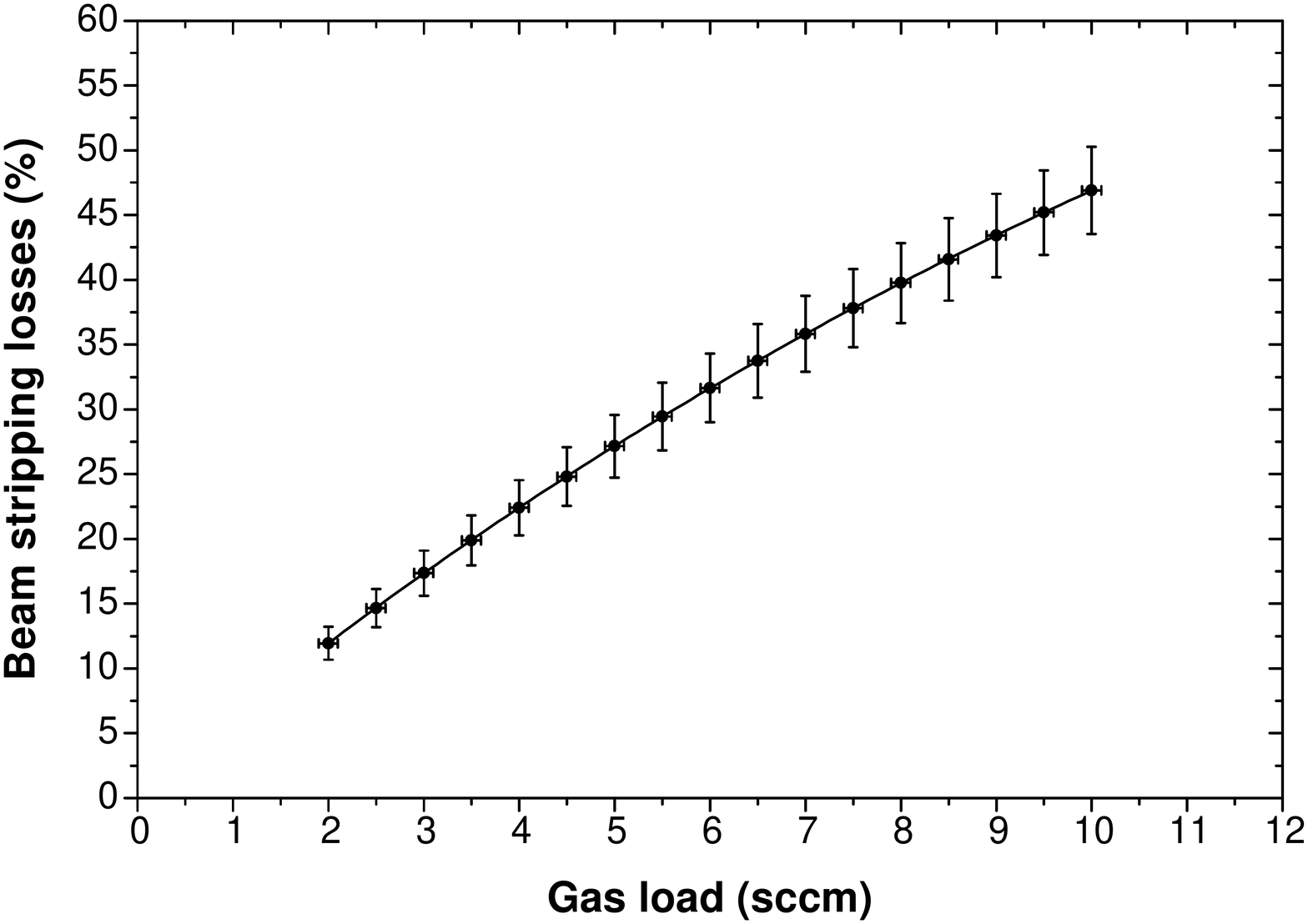} \label{fig:Losses-gasLoad-H2} } $\;$
	\subfigure[]
	{\includegraphics [width=8.6cm] {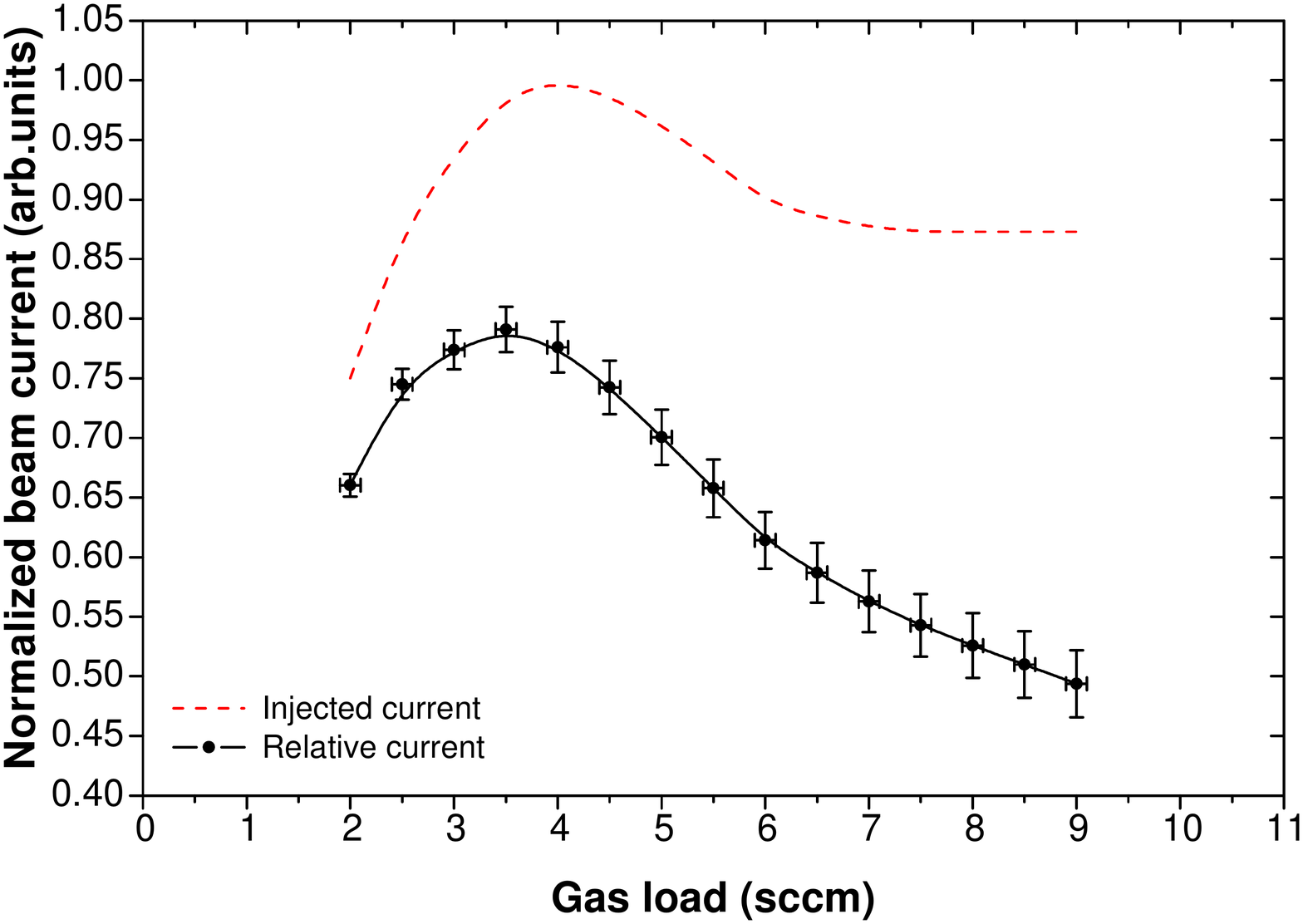} \label{fig:gasLoad-RelativeCurrent} }
	\caption{(a) Stripping losses rate as a function of the gas load of the chimney. (b) Relative final beam current from stripping losses simulations considering the experimental results of the ion source extraction measurements. To facilitate the data analysis the red dashed line represents the normalized injected current over the maximum from the experimental measurements~\cite{calvo-2}.}
	\label{fig:Losses-gasLoad}
\end{figure}

The stripping loss rate as a function of the gas load has been obtained (Fig.~\ref{fig:Losses-gasLoad-H2}), considering the actual flow range provided by the gas supply system of the ion source ($0-10$~sccm). The gas load is of great relevance in the fulfillment of the cyclotron beam requirements, because it influences directly the final beam current through two associated effects. On the one hand, the gas load influences on the injected beam. In accordance with the experimental results obtained from the measurements of the \amit ion source in a test bench facility~\cite{calvo-2}, the injected beam current presents a non-linear behavior with the gas flow rate. The production and the neutralization of negative ions take place simultaneously by different reaction mechanisms inside the plasma source~\cite{kuo}. Then, a balance between the gas ionization rate inside the chimney and the survival of H$^-$ ions led to a maximum extracted beam current at $4$~sccm of gas load for different operating conditions of the ion source. In spite of the fact that these experimental results were obtained under DC extraction conditions, the behavior of the extracted current from the source must be proportional in RF conditions, and therefore, it is acceptable to use these measurements to quantify the expected current in the cyclotron. On the other hand, higher gas load implies higher stripping losses by the increment of the pressure level. The increase of the beam losses can even reduce the final beam current below the minimum needed for the radioisotope production. Therefore, the optimum solution will require a balance between both effects. The results of the optimum gas flow rate to maximize the injected current in combination with the limited beam stripping losses allow to establish a gas load of $3.5$~sccm to achieve a maximized final beam current (Fig.~\ref{fig:gasLoad-RelativeCurrent}), slightly lower than the optimum when only ion source current is considered.

\begin{figure}[!b]
    \centering
    \subfigure[]
    {\includegraphics [width=8.6cm] {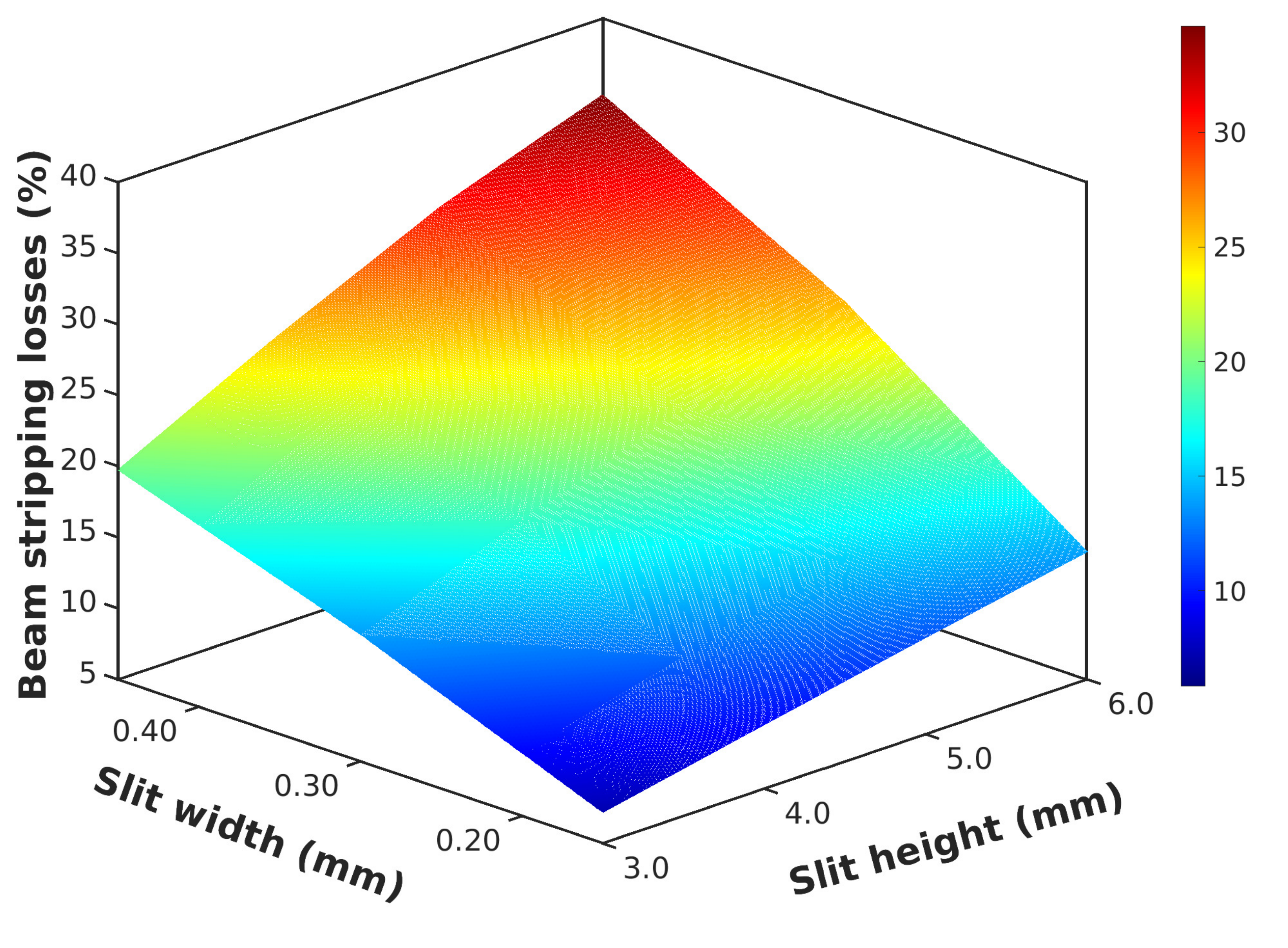} \label{fig:Losses-ChimneyArea}}
    \subfigure[]
    {\includegraphics [width=8.6cm] {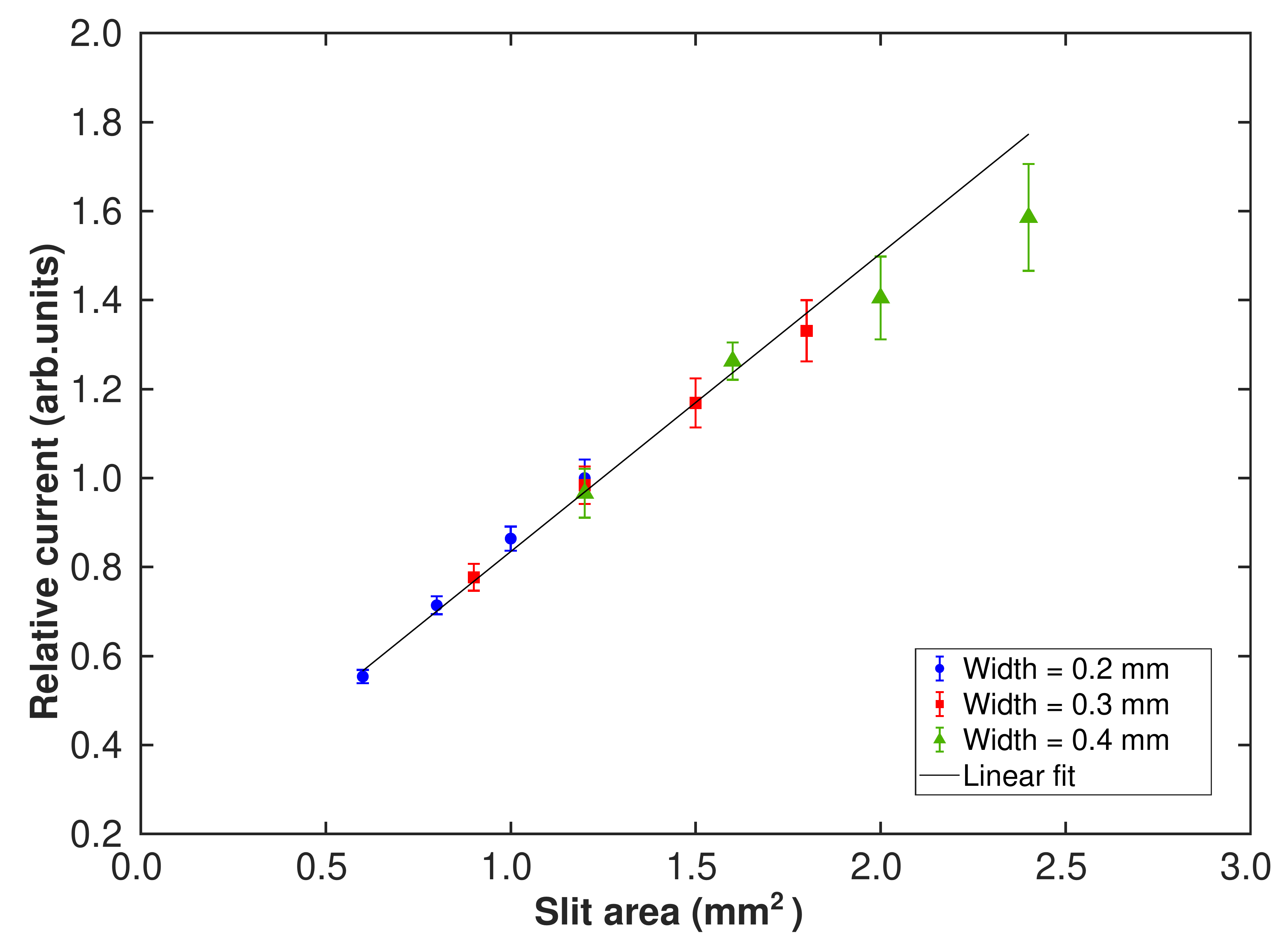} \label{fig:RelativeCurrent-ChimneyArea}}
	\caption{(a) Beam stripping losses rate as a function of the slit height and width. The color surface mesh is interpolated from the results of the simulations. (b) Normalized relative current as a function of the slit area. The values are normalized over the expected current from the nominal slit ($6\!\times\!0.2$~mm$^2$). The solid red line is a linear fit to the data weighted over the stochastic error associated with the stripping losses. The results are grouped by color and symbol according to the slit width, with each point representing a different slit height.}
	\label{fig:Losses-Area}
\end{figure}

In addition to the gas load, the stripping losses can be optimized by the adjustment of the chimney slit size, which modifies the conductance. From the designed nominal size of $6$~mm height and $2$~mm width, different slit areas have been considered from $0.6$~mm$^2$ to $2.4$~mm$^2$, modifying both the height (between $6$, $5$, $4$ and $3$~mm), and the width (between $0.2$, $0.3$ and $0.4$~mm). A linear increase of the pressure level in the vacuum chamber with the slit size has been considered. Multiparticle simulations have been performed to evaluate its effect on the beam stripping losses along the cyclotron, observing, as expected, a linear increase with the size of the slit (Fig.~\ref{fig:Losses-ChimneyArea}) both for changes in height or width.

The slit size also has an impact on the injected current, the beam size and the associated losses due to axial deviations. First, considering a linear relationship between the injected current and slit area, a net increase on the final beam current will be expected as the area is modified in the considered range, in spite of the rise of beam stripping with the residual gas. Moreover, the wider the beam width, the greater the losses in the central region due to particle collisions with the puller. Fig.~\ref{fig:RelativeCurrent-ChimneyArea} represents the relative beam current at the end of the cyclotron in relation to the expected current from the nominal slit. The significant increase in the final current is clearly noticeable. Therefore, it can be concluded that a larger slit width than the nominal will be an advantageous enhancement on the final beam current. A deviation from the linear trend is observed due to the increase of losses in the central region due to impact with the puller. Lastly, a slight reduction in the height of the slit entails the complete axial beam transmission, reached for a slit height of $\leq\!5.5$~mm. Larger slit height leads to particle collisions with the top and bottom of the electrode in the medium-high energy range, that must be avoided to prevent the activation of the accelerator. Thus, the maximum slit height must be set below this value in order to avoid these beam losses.

Based on the results obtained, the analysis of the beam losses for different slit sizes allows to establish the range of dimensions that provide an optimization of the final current without an excessive rise of the stripping losses. It should be mentioned that optimizing the slit dimensions shall also consider the effect on the plasma meniscus and the extraction electric field modification. These effects will be analyzed in future studies.

The results obtained in this section will provide a starting point during the commissioning of the cyclotron to validate the simulations against experimental data, balancing all the related parameters to optimize the final beam current.


\section{Conclusions}

Beam stripping interactions with residual gas and electromagnetic fields (Lorentz stripping) have been studied in detail in order to analyze the effect on the cyclotron beam transmission through the implementation of these physical interactions into the beam dynamics code \opal. These interactions constitute one of the main drawbacks in H$^-$ compact cyclotrons with internal ion source, due to their influence on the beam transmission, the high potential for damaging the cyclotron in-vessel components and the associated radioactive environment. The importance of the gas stripping has been corroborated by the application to the \amit cyclotron. Thanks to this assessment and optimization of the ion source parameters and the vacuum system has been obtained. Firstly, the control of the gas load has been studied taking into account previous experimental measurements of the current extracted from the ion source. The balance of beam production at the ion source and the increase of stripping losses has been evaluated. Secondly, we assessed the effects of the ion source slit size and the injected beam current on the stripping losses and on the central region and the beam axial deviations. Bearing that in mind, stripping interactions have to be considered during the beam dynamics simulations of future compact H$^-$ cyclotrons, as one of the key effects for the optimization of the nominal parameters. The \opal code has been upgraded and validated as a powerful tool for this use.


\section*{Acknowledgement}

The authors thank PSI and OPAL team for supporting this research. All simulations were performed using the HPC resources provided by the computing facilities of Extremadura Research Centre for Advanced Technologies (CETA-CIEMAT), funded by the European Regional Development Fund (ERDF). CETA-CIEMAT belongs to CIEMAT and the Government of Spain. This work was partially supported by the Spanish Ministry of Economy and Competitiveness under project FPA2016-78987-P.


\bibliography{bibfile}

\end{document}